\documentclass[12pt,preprint]{mnras}
\usepackage{amsmath}
\usepackage{graphicx}	% Including figure files
\usepackage{amssymb}	% Extra maths symbols

\title[AIC of magnetic white dwarfs in binaries]{The Magnetized
White Dwarf + Helium star Binary Evolution with Accretion-induced Collapse}

\author[I. Ablimit]
{Iminhaji Ablimit$^{1,2}$\thanks{E-mail: \href{mailto:iminhaji@nao.cas.cn}{iminhaji@nao.cas.cn}(I.A)}
\\
$^{1}$Key Laboratory for Optical Astronomy, National Astronomical Observatories,
Chinese Academy of Sciences, Beijing 100012, China\\
$^{2}$Department of Astronomy, Kyoto University, Kitashirakawa-Oiwake-cho, Sakyo-ku, Kyoto 606-8502, Japan}

\begin{document}
\label{firstpage}
\pagerange{\pageref{firstpage}--\pageref{lastpage}}
\maketitle

\begin{abstract}

Accretion-induced collapse (AIC) from oxygen/neon/magnesium composition white dwarf (ONeMg WD) + stripped helium (He) star binaries
is one promising channel to form peculiar neutron star objects.
It has been discussed that
the WD's magnetic field may alter the accretion phase in the WD binary evolution.
By considering non-magnetic and sufficiently magnetized WDs, we investigate the evolution of ONeMg WD + He star binaries with detailed stellar evolution and
binary population synthesis simulations. The role of the magnetically confined accretion in the possible formation pathway for like millisecond pulsars (MSPs) and magnetars is also studied. Comparing with the case of spherically
symmetric accretion, the mass accumulation
efficiency of the WDs is enhanced at low mass transfer rate under the magnetic confinement model. The initial parameter space
of the potential AIC progenitor systems moves toward shorter orbital period and lower donor mass (but not so significantly) due to the effect of the magnetic confinement.
This also allows final MSPs to have lower-mass WD companions and shorter orbital periods. There is no significant difference between the Galactic birthrates of the AIC derived with and without the magnetic confinement, which implies that the magnetic field of the WD does not dramatically change the number of ONeMg WD + He star binaries which can produce AIC.
It is worth noting that these conclusions can be applied for the CO (carbon/oxgen) WD
+ He star binaries as progenitors of type Ia supernovae, because the accretion phases of ONeMg WDs and CO WDs are similar.
The Galactic rate of magnetars possibly formed via AIC of highly magnetized WDs is $0.34\times10^{-4}\,{\rm yr}^{-1}$.

\end{abstract}


%%%%%%%%%%%%%%%%%%%%%%%%%%%%%%%%%%%%%%%%%%%%%%%%%%

%%%%%%%%%%%%%%%%% BODY OF PAPER %%%%%%%%%%%%%%%%%%

\section{Introduction}

 Binaries containing white dwarfs (WDs) and stripped (naked) helium (He) stars
are promising multimessenger sources for the upcoming
electromagnetic and gravitational wave detectors, because they can evolve to produce
a number of astrophysical transient phenomena and peculiar
objects such as: helium novae, type Ia supernovae (SNe Ia), AIC, AM Canum Venaticorum (AM CVn) stars, millisecond pulsars (MSPs), and pulsar binaries with WD
companions (Nomoto 1982; Iben \& Tutukov 1991;
Woosley \& Weaver 1994; Bildsten et al. 2007; Deloye et al. 2007; Kato et al. 2008; Stroeer \& Nelemans 2009;
Woosley \& Kasen 2011; Townsley et al. 2012;
Brooks et al. 2017).

Observationally, most MSPs are found in binaries (e.g., Manchester et al. 2005),
while all known magnetars up to date are isolated and strongly magnetized NSs (e.g., Olausen \& Kaspi 2014).
It is possible to have both MSPs and magnetars born in binary systems.
The AIC of a ONeMg WD has been proposed as one important way to form proto-neutron stars (NS) with
millisecond spin periods (e.g., Bhattacharya \& van den Heuvel 1991; Freire \& Tauris 2014; Ablimit 2019), and the NS through the AIC with
amplified magnetic field as magnetar (e.g., Duncan \& Thompson 1992).

In the evolution pathway of the AIC, an ONeMg WD accretes hydrogen-rich or helium rich materials from its donor star and grows
in mass to the Chandrasekhar limit, and the following process is the electron capturing which lets the WD collapse to
produce a fast spinning NS (e.g., Nomoto \& Kondo 1991; Ablimit \& Li 2015).
The possibility of direct formation of MSPs by the AIC has been discussed in previous works (e.g., Chanmugam \& Brecher 1987;
Michel 1987; Kulkarni \& Narayan 1988; Yoon \& Langer 2005).
Binary population synthesis (BPS) results of Hurley et al. (2010) have presented
the significant contributions of the AIC scenario to
the formation of MSPs. Ruiter et al.(2019) investigated different evolutionary pathways of AIC to form NSs, and also showed their important contributions (see also Wang \& Liu 2020 for a review). However, the possible contributions of magnetism in ONeMg WD + He star binaries to form peculiar NS populations are not investigated yet, and we study the role of the magnetism in the ONeMg WD + He star binary stellar evolution to yield AIC in this work.

The number of observed
magnetic WDs is increasing, and they have been found in cataclysmic variables (e.g., Ferrario et al. 2015), symbiotic binaries and super-soft X-ray
sources (Kahabka 1995; Sokoloski \& Bildsten 1999; Osborne et al. 2001). Highly magnetized WDs like polars have a different accretion mode (i.e. two-pole accreting) compared to non-magnetic or weakly magnetized WDs (e.g., Schwope et al. 2021).
However, not just the AIC itself but also the highly magnetized WD binary evolution as the progenitor of AIC or SNe Ia still lack of direct observational evidence (e.g., Ablimit \& Li 2015, 2019; Ruiter et al. 2019), and polar systems observed to date can not generate AIC/ SNe Ia, because they have very low mass donors or/and low mass WDs needed for producing SNe Ia or AIC (see below sections for more information).
Theoretically, it has been studied that the magnetic field of the WD may play a key role
in the evolution of CO (carbon/oxgen) WD binaries,
and the CO WD even with a low mass companion can grow in mass and reach the Chandrasekhar limit mass to explode
as a SN Ia if magnetic confinement is achieved (e.g., Ablimit \& Maeda
2019a,b). The magnetic confinement model has been proposed to have the crucial effect on accreting and nuclear burning processes of the WD, and it is related to the magnetic field strength of the WD, mass transfer rate (mass of the donor) and mass of the WD  (e.g., Livio 1983; Ablimit \& Maeda 2019a; Ablimit 2019). Thus, in this study, we investigate the influence of the magnetic confinement model in the evolution of ONeMg WD + He star binaries. We also point out the same role of the magnetic field in the evolution of CO WD + He star binaries to SNe Ia, because the mass transfer/accretion phases in the CO WD binaries are the same as that of ONeMg WD binaries (see Hillman 2021).

In this work, we present detailed numerical calculations of the ONeMg WD $+$ He star binaries by considering
non-magnetic and magnetic cases in the pre-AIC evolution (e.g., Ablimit 2019).
The main goal of the work is to show the possible role of the magnetic confinement model in the WD $+$ He star binary evolution.
If the ONeMg WD in the AIC channel is strongly magnetized, the newborn NS may inherit the large magnetic field through
 conservation of magnetic flux, and the newborn NS could be a magnetar (see section 3 for more discussions).
In this way, we study the formation of MSPs (or pulsars) and magnetars in WD binary systems with He star companions,
and we perform BSP simulations based on Ablimit \& Maeda (2019b) for deriving birth rates.
  In \S 2,
we describe the models for the evolution of WD $+$ naked (or stripped) He star binaries, BPS simulations for rates, and discuss results.
We also briefly discuss that our results/conclusions can be used to evolution of CO WD + He star binaries to SNe Ia in \S 2.
We introduce the post-AIC evolution pathway, and also present (millisecond)pulsar-WD systems left from post-AIC in \S 3.
The paper is closed in \S 4 with the conclusions.

%\section{The calculation model}

\section{The Evolution of WD $+$ stripped He star binaries (pre-AIC) without and with the Magnetic Confinement Model}
\label{sec:model}

\subsection{Detailed Binary Evolution Model: Formation of MSPs and Magnetars}

In this section, the WD + He star binary evolution is simulated by using the version 10389
of Modules for Experiments in Stellar Astrophysics (MESA; Paxton et al. 2015),
and considering the magnetic confinement model.  He
abundance Y = 0.98 and metallicity Z = 0.02 are taken for the naked non-degenerate helium donor star.
WDs with a initial mass of 1.2 $M_\odot$ are used as the accreting primaries in those binaries.
Interestingly, a recent study of Hillman (2021) found that the evolution of the ONeMg WD model including the accretion phase has results that are
almost identical to those of the CO WD model. Thus,
after the WDs reach the Chandrasekhar limit mass
($M_{\rm Ch} =$1.38$\rm M_\odot$), it can be AIC to form NSs or SNe Ia according to the ONeMg core or CO core of the WDs, respectively. In this work, we focus on the ONeMg WD binaries, and we also calculate
post-AIC evolutions with the MESA code.
In the code, the prescription of Eggleton (1983) is used
for calculating the effective Roche-lobe (RL) radius
of the donor star ($R_{\rm RL, d}$),
\begin{equation}
 R_{\rm RL, d} =(\frac{0.49q^{2/3}}{0.6q^{2/3} + {\rm ln}(1+q^{1/3})})a,
\end{equation}
where $q = M_{\rm donor}/M_{\rm accretor}$ and $a$ is the orbital separation. The Ritter scheme (Ritter 1988) is adopted for the mass transfer via RL overflow (RLOF),
\begin{equation}
{\dot{M}} = -\dot{M}_{0}\,{\rm exp}(\frac{ R_{\rm d} - R_{\rm RL, d}}{H_{\rm P,d}/\gamma(q_2)}),
\end{equation}
where $\dot{M}$ is the RLOF mass transfer rate, $ R_{\rm d}$ is the radius of the donor,
and $H_{\rm P,d}$ is the pressure scale height at the photosphere of the
donor. Calculations of $\dot{M}_{0}$ is
\begin{equation}
\dot{M}_{0} = \frac{2\pi}{\rm exp(1/2)}F(q_2)\frac{R^3_{\rm RL, d}}{GM_d}{(\frac{\kappa_{\rm B} T_{\rm eff}}{m_{\rm p} \mu_{\rm ph}})^{3/2}}\rho_{\rm ph},
\end{equation}
where $\kappa_{\rm B}$ is Boltzmann constant, $m_{\rm p}$ is the proton mass, and $T_{\rm eff}$ is the effective temperature of
the donor. $\mu_{\rm ph}$ and $\rho_{\rm ph}$ are the mean molecular weight and density at its photosphere. The two fitting functions $F({q_2})$ and $\gamma(q_2)$ ($q_2 = M_{\rm accretor}/M_{\rm donor}$) are,
\begin{equation}
\begin{array}{ll}
F(q_2) =
1.23 + 0.5{{\rm log}(q_2)} & \textrm{$0.5\lesssim q_2 \lesssim 10$},
\end{array}
\end{equation}
and,
\begin{equation}
\gamma(q_2) = \left\{ \begin{array}{l}
0.954 + 0.025{{\rm log}(q_2)} - 0.038{({\rm log}(q_2))^2} \\
0.04\lesssim q_2 \leqslant 10.954 + 0.039{{\rm log}(q_2)} + 0.114{({\rm log}(q_2))^2}  \\
1\leqslant q_2 \lesssim 20
\end{array} \right.
\end{equation}
Validation of $F({q_2})$ and $\gamma(q_2)$ are same as in the MESA code (Paxton et al. 2015; see also Ritter 1988).
Other physical issues in the evolutions are same as typical ones introduced in the instrumental papers of MESA (e.g., Paxton et al. 2015).
The mass retention efficiency is one of the important things which causes different results
in the accreting WD. The calculation methods of Kato \& Hachisu (2004) is used
for the mass accumulation efficiency of helium, then the mass growth rate of the WD is
$\dot{M}_{\rm{WD}} = \eta_{\rm He} {\dot{M}}$,
where $\eta_{\rm He}$ is efficiency of helium burning. Because the maximum accretion rate calculated in
Kato \& Hachisu (2004) is $10^{-5.8}\,\rm{M_\odot\,yr^{-1}}$, we set a constant accretion rate of $10^{-6}\,\rm{M_\odot\,yr^{-1}}$ if the accretion rate exceeds $10^{-6}\,\rm{M_\odot\,yr^{-1}}$ (see also Brooks et al. 2017).
The comparison of different retention efficiency models and critical mass transfer regions
for stable He burning can affect evolution results (see Brooks et al.(2016) for more discussions).
Brooks et al.(2017) find that the C/O ashes from the steady
helium burning shell unstably ignite in a shell flash, but that
these carbon burning episodes do not significantly interrupt the
growth of the WD.
The angular momentum loss caused
by mass loss and gravitational wave radiation (GR, Landau \& Lifshitz 1975) is included.
The magnetic confinement model in the magnetized WD + He star binary evolutions are presented as below.

Comparing to the spherically symmetric accretion onto non-magnetic WDs, the accretion disk formation in a sufficiently magnetized WD binary would be partially
disrupted or prevented by the WD's magnetic field (e.g., Cropper 1990; Frank et al. 2002). The accretion phase and emissions in the magnetized WD can be different due to the strong magnetic field (e.g., Fabian et al. 1977; Angel 1978; King \& Shaviv 1984; Hameury et al. 1986; Wickramasinghe \& Ferrario 2000).  When the RLOF mass transfer occurs, the stream of matter
from the donor is captured by the magnetic field of the WD, then it follows the magnetic
lines and falls down onto the magnetic poles of the WD as an accretion column (e.g., King 1993).

For accreting highly magnetized WDs with dipolar magnetic fields, the magnetic pressure (Frank et al. 2002) increases rapidly as the accreted material approaches the WD's surface, at the magnetospheric radius, the magnetic pressure could exceed the ram pressure of the accreting matter, and the motion of the accreting matter starts to be controlled by the magnetic fields.
According to King (1993), Frank et al.(2002) and  Wickramasinghe (2014), the minimum field of the WD for the stream/comfined accretion of the helium-rich material can be written as,

\begin{multline}
\rm{B} \geq \rm{B_c} \simeq 7.2\times10^{5} (\frac{\dot{M}}{10^{-10}\,\rm{M_\odot\,yr^{-1}}})^{1/2} (\frac{\rm{P_{orb}}}{\rm day})^{5/6}\\
(\frac{\rm{M_{WD}}}{\rm M_\odot})^{7/6} (\frac{\rm{R_{WD}}}{10^9\,\rm cm})^{-3}\,\, {\rm Gauss},
\end{multline}
where $\dot{M}$ is the RLOF mass transfer. $\rm{P_{orb}}$ is the orbital period of the WD binary.
For the mass ($M_{\rm WD}$) -- radius ($\rm{R_{WD}}$) relation of the WD, the equation given in Nauenberg (1972) is used.
With this minimum condition, it is assumed that the magnetic pressure can be larger than the pressure of the helium-rich matter (see Shen \& Bildsten (2009) for He burning pressure).
The magnetic confinement needs the magnetic field which satisfies the condition given in the equation (6).
In the code, we set the magnetic field strength of the WD with a fixed value, and compare it with the calculated value from the right side of equation (6). If it satisfies condition (6), then the confined/stream mass accretion is achieved and the mass accretion phase is simulated as described below.
We have run a large number of WD binary evolutions by testing non-, intermediate- and high-magnetic field strengths (B) of the WDs.
Here, the sufficiently magnetized WD with the fixed initial magnetic field strength of
$2.5\times10^7$ G is taken to show the main results.

%\begin{multline}
%\rm{B} \geq 9.3\times10^{7} (\frac{\rm{R_{WD}}}{5\times10^8\,\rm cm})(\frac{\rm{P_b}}{5\times10^{19}\,\rm{dyne\,cm^{-2}}})^{7/10}\\
%(\frac{\rm{M_{WD}}}{\rm M_\odot})^{-1/2}(\frac{\dot{M}}{10^{-10}\,\rm{M_\odot\,yr^{-1}}})^{-1/2}\, Gauss,
%\end{multline}

Because of the confinement induced by the magnetic field,
the WD can have the isotropic pole-mass transfer rate ($\dot{M}_{\rm{p}}$) as (e.g., Ablimit \& Maeda 2019a),

\begin{equation}
\dot{M}_{\rm{p}} =  \frac{\rm S}{\Delta \rm{S}} {\dot{M}},
\end{equation}
where $\dot{M}$ is the RLOF
mass transfer rate, $\rm S$ is the surface area of the WD, and $\Delta \rm{S}$ is that of the dipole regions. The stream-like mass accretion phase can be modeled with $\dot{M}_{\rm{p}}$ for the confinement model. The surface fraction is $\frac{\rm S}{\Delta \rm{S}}=\frac{\rm{2 R_{m}}}{\rm{R_{WD}} \rm{cos^2{\theta}}}$,
where $\rm{R_{m}}$ is the magnetosphere radius related with Alfv$\acute{\rm e}$n radius of the magnetic WD
(see Ghosh \& Lamb (1979) and Frank et al. (2002) for the calculation and discussions of $\rm{R_{m}}$; Note that $\rm{R_{m}}$ depends on the accretion and magnetic field evolution.
$\theta$ is the angle between the rotation axis and magnetic field axis.
We take $\theta = 0$ for the simulations in this work.)
Here with the magnetic confinement model, the He burning efficiency is determined
by $\dot{M}_{\rm{p}}$ (Ablimit \& Maeda 2019a).

In Figure 1, we demonstrate one typical WD + He star binary evolution under the magnetic confinement model.
Without the effect of the WD's strong magnetic field, the WD ($M_{\rm WD,i}=1.2 M_\odot$) cannot grow in mass
to the Chandrasekhar limit mass due to the low mass transfer rate from a naked He star
with a initial mass of 0.8 $M_\odot$, especially almost no mass growth during late RLOF mass transfer phase.
The strong magnetic field confines the accreted matter in
the polar cap regions, and alter the accretion phase (Figure 1). The main difference of these two cases with the same RLOF mass transfer from the donor is that the accretion rate per unit area on the WD due to the magnetic confinement is higher than that in the spherical accretion case. The WD accretes matter with a constant rate of $10^{-6}\,\rm{M_\odot\,yr^{-1}}$ (green line in the figure) when the rate $>10^{-6}\,\rm{M_\odot\,yr^{-1}}$.
The WD is growing in mass smoothly while the He donor star is reducing its mass (lower panel of Figure 1).
The WD may not accrete all transferred mass due to the high mass transfer rate (red line), some of the matter may be lost from the binary as the optically thick wind (see Hachisu et al. 1996),
and the mass loss will take the angular momentum with it.
This typical evolution (Figure 1) shows main influence of the magnetic confinement model in the ONeMg WD + He star binary.

Based on the example, the upper panel of Figure 2 shows the initial parameter
space of ONeMg WD + He star binaries with $M_{\rm WD,i}=1.2 M_\odot$ for non-magnetic and magnetic confinement cases. We made the calculation with initial He star masses $M_{\rm He,i}$ from 0.5 $M_\odot$ to 3.0 $M_\odot$ by steps of 0.15 $M_\odot$, and the initial orbital periods $P_{\rm orb,i}$ logarithmically from -1.5 to 2.8 (in days) by steps of 0.15.
We find that the initial parameter spaces with the magnetic confinement case which lead to AIC move toward shorter orbital periods ($-1.15\leq {\rm{log_{10}}(P_{\rm orb,i})}\leq 1.2$ days) and lower He donor
masses ($0.6 < M_{\rm He,i}\leq 2.6 M_\odot$).
Results of this work and results of Liu et al. (2018) have some differences (our parameter space is smaller) which is
caused by different prescriptions in the different stellar evolution codes, such as the different RLOF mass transfer scheme.
Tauris et al. (2013) derived a smaller parameter space of
ONe WD + He star systems to generate AIC compare to our result. Two main reasons may cause this discrepancy.
First, Tauris et al. (2013) adopted $M_{\rm Ch} =
1.48M_\odot$ for rigidly rotating WDs, while we used $M_{\rm Ch} =
1.38M_\odot$.
Second, Tauris et al. (2013) assumed and adopted one criterion for the binary evolution to enter
the common envelope phase, which is $M_{\rm He}>M_{\rm Edd,WD}$.
However, $M_{\rm Edd,WD}$ adopted in their work is for H-accreting WDs, but too small for He-accreting
WDs as the opacity of helium is lower than hydrogen. Corresponding to initial parameter space,
the donor masses and orbital periods just after
AIC shift to lower regions compared to the non-magnetic case (lower panel of Figure 2).
Newborn NSs could be MSPs (note that they may possibly be pulsars, it depends on evolution) or Magnetars with He star companions, and they
may act like ultraluminous X-ray sources if the He donor starts RLOF mass transfer again (e.g., Abdusalam et al. 2020).
Non-magnetized material accretion onto the WD may decay the magnetic field of the WD (e.g., Cumming 2002),
 we assume that the WD still could have strong enough field when it begins to collapse for the case of magnetar formation.

\subsection{Birth Rates}

In this subsection, we simulate $3\times10^7$ binary systems from the initial stage of zero-age main sequence
by using Monte-Carlo binary population synthesis code (Hurley et al. 2002; Ablimit et al. 2016; Ablimit \& Maeda 2018; Ablimit et al. 2021)
in order to form and evolve the ONeMg WD + He star binaries.
ONeMG WD +  He star binaries can be formed by experiencing at least one common envelope (CE) phase
 when the initial binary configuration leads
to unstable mass transfer, for example: a binary evolves into the first CE phase when the primary star
(the more massive one) fills its RL during its asymptotic giant branch evolutionary stage, and after the successful
CE ejection, the system enters the second CE evolution when the secondary fills its RL and begins unstable
rapid mass transfer (e.g., Iben \& Tutukov 1994; see Wang \& Liu (2020) for other routes).
The two CE ejections leave a ONeMg WD + He star binary in
a short orbit due to a large fraction of the initial angular momentum taken by the mass loss.

The initial input parameters of the primordial binaries taken in this work are:
the initial mass function of Kroupa et al. (1993) is
adopted for the masses of the primary stars; a flat distribution (e.g., Ferrario 2012) between 0 and 1 is used for the distribution of the mass
ratio of the secondary to the primary, and secondary mass can be determined by the mass ratio. All stars are assumed to be initially in
binaries, and the initial metallicity of stars are set as the solar metallicity. The distribution of the orbital separations is assumed
to be flat in logarithm (e.g., Connelley et al. 2008). The thermal distribution between 0 and 1 is used for initial eccentricity
distribution in the calculation.
Ablimit et al. (2016), Ablimit \& Maeda (2018) and Ablimit (2021) have discussed more details of our BPS code. The CE
ejection efficiency ($\alpha_{\rm CE}$) is taken to be 1.0 in this paper.
The more recent and physically motivated treatment for the binding energy parameter ($\lambda$) is adopted in the CE evolution (Wang et al. 2016; Klencki et al. 2021).
The main purpose of this paper is to show the potential contributions of the highly magnetized WDs with He stars to the AIC.

By following the BPS simulations of Magnetized WD binaries in Ablimit \& Maeda (2019b),
two models (Model non-mag and Model mag) are calculated in this BPS calculations.
In Model non-mag, all formed WD + He star binaries are assumed to
be non-magnetic WD binaries, while all WDs are assumed to be born with the
high magnetic field in the WD/He star binaries in Model mag. Because the main aim of this BPS
calculation is to demonstrate the possible contribution of magnetic confinement model for MSPs and
magnetars formed through AIC, we adopt the same typical initial
conditions for non-mag and mag models, except the condition of the magnetic field.

The WD gains angular momentum with the accreted matter during the pre-AIC process to reach
the Chandrasekhar limit ($M_{\rm Ch}$), thus it can be spun up and it is assumed that the accreting and fast rotating WD collapses to
form a MSP. If the WD in the AIC channel is strongly magnetized, the
newborn MSP (might be a pulsar) may inherit the large magnetic field through amplified
magnetic field (e.g.,
Duncan \& Thompson 1992). Indeed, recent observational result of FRB 20200120E (Kirsten et al. 2021) indicate that
the source contains a young and highly magnetized NS, which may be formed via AIC of a magnetic WD. Therefore, it is hypothesized in this work
that a magnetar could be the result of the AIC formation channel from a
highly magnetized WD system. It is very difficult to obtain (at least so far) the real fraction of highly magnetized ONeMg or CO WDs among all WD binaries, thus here the fraction of sufficiently magnetized WDs
among the WD population is assumed as 18\% to derive the possible birthrate of the magnetars based on the available fraction of polars among observed CVs (about 15\%-18\%: Ferrario et al. 2015; see also Kepler et al. 2013, 2015). Note that the assumed fraction may be overestimated, and we expect more constraints on it from future observations. In
the calculation of the Galactic birthrate (e.g., Ablimit et al. 2016; Abdusalam et al. 2020), a constant star formation rate (SFR) over the past 13.7 Gyr is considered,
and $5\,\rm{M_\odot\,yr^{-1}}$ is taken for the SFR according to Willems \& Kolb (2004).

The Galactic birthrate of AIC as a function of time
with a constant SFR of $5\,M_\odot\,{\rm yr}^{-1}$ is given in Figure 3. The solid black
line and red dashed line show the rates for cases of with and without the magnetic confinement model. The predicted birthrates
of AIC in non- and -magnetic models are $1.4\times10^{-4}\,{\rm yr}^{-1}$ and $1.9\times10^{-4}\,{\rm yr}^{-1}$, respectively.
The comparison between the birthrates of AIC from non-magnetic and magnetic cases indicates that the magnetic confinement model does not significantly change the contribution
of ONeMg WD + He star binary channel to AIC rate and formation rate of NS binaries.
If we take the observed inferred fraction of magnetic and non magnetic WDs into consideration,
birthrates of MSPs and magnetars can be $1.15\times10^{-4}\,{\rm yr}^{-1}$ and $0.34\times10^{-4}\,{\rm yr}^{-1}$,
and whole MSP population formed through AIC become $1.49\times10^{-4}\,{\rm yr}^{-1}$. Our rate has good agreement with the rates derived by previous works (e.g., Hurley et al. 2010; Ruiter et al. 2019; Wang \& Liu 2020).
The MSPs formed via AIC scenario is especially favored
for the MSPs in globular clusters, because they have characteristic ages (from tens Myr to few hundreds Myr)
significantly less than the ages of the clusters.
This result also shows that this may be the solution for the birth rate problem between Galactic low-mass X-ray binaries (LMXBs)
and MSPs (e.g., Kulkarni \& Narayan 1988; Levin et al. 2013; Ferrario \& Wickramasinghe 2007).

\section{The post-AIC Evolution: Formation of MSP - WD Systems}

We assume that the WD collapses to be an NS with a gravitational mass of
1.25 $M_\odot$ when ${M}_{\rm WD} = M_{\rm Ch}$. A mass of 0.13 $M_\odot$ from the WD converts
into released gravitational binding energy during the AIC process (see also Ablimit \& Li 2015).
The orbital separation becomes wider due to the sudden mass-loss. According to the prescription
of angular momentum conservation, the relationship
between the orbital separations just before ($ a_0$) and after ($ a$) the collapse (Verbunt et al. 1990) is calculated as,

\begin{equation}
\frac{a}{a_0} =  \frac{M_{\rm WD} + M_2}{M_{\rm NS} + M_2},
\end{equation}
where $M_{\rm NS}$ and $M_2$ are the NS and He star masses. The newborn NS through AIC would have a very
small kick (Hurley et al.2010; Tauris et al.2013), thus we do not include effect of the kick. The eccentric
orbit by the mass loss during the AIC could be efficiently circularized by the tidal effect.

After a magnetar or MSP binary with the He star formed, the RLOF mass transfer occurs again
due to the expansion of the donor. Accreted material would decay the magnetic field strength of the NS (e.g., van den Heuvel
\& Bitzaraki 1995), thus the strong field of the magnetar can decay with increase in the accreted mass,
and magnetar would become a normal MSP.
Here, the spin evolution, magnetic decay due to accretion, and effect of the magnetic
field of the NS during the accretion process are out of the scope of this work.
An accretion disk will
be formed when the transferred matter surrounds the NS.
To form a stable accretion disk, the mass transfer rate should be higher
than the critical value given as (Dubus et al. 1999),

\begin{equation}
\dot{M}_{\rm c} =  3.2\times10^{-9}{(\frac{M_{\rm NS}}{1.4\rm M_\odot})^{0.5}}{(\frac{M_2}{1.0\rm M_\odot})^{-0.2}}
{(\frac{P_{\rm orb}}{1.0\,\rm d})^{1.4}}\,\rm M_\odot \rm{yr}^{-1},
\end{equation}
where $P_{\rm orb}$ is the orbital period. The mass transfer process is nonconservative
and Eddington limited ($\dot{M}_{\rm Edd}$), the NS accretion is as following,

\begin{equation}
\dot{M}_{\rm NS} =  \rm{min}(-\beta \dot{M}, \dot{M}_{\rm Edd}),
\end{equation}
where $\dot{M}$ is the RLOF mass transfer from the donor to NS. The mass accretion and mass loss processes are still not well constrained.
Thus, we take a typical value of $\beta = 0.5$, and the excess matter is assumed to be expelled
by the NS in the isotropic wind carrying the specific angular momentum. The
newborn MSPs can spin up\footnote{An NS with a typical mass can spin up to 50 ms from rest if it accretes only a small
amount of mass ($10^{-3} {M_\odot}$) (see the equations and discussions
of Tauris et al. (2012) for more details).} faster by accreting mass, and a magnetar also can spin faster but
its magnetic field strength decays due to the accretion. Considering the spun-up and magnetic field decay
due to mass accretion, all final systems are taken as MSP - WD systems.
For the angular momentum loss (AML) during the post-AIC evolution, it can be expressed as,
\begin{equation}
\dot{J} =  \dot{J}_{\rm GR} + \dot{J}_{\rm ML},
\end{equation}
where is $\dot{J}_{\rm GR}$ and $\dot{J}_{\rm ML}$ are the AML due
to the GR and mass loss, respectively. Here, note that the evaporative wind which may be caused by the pulsar's rotationally powered
luminosity is not considered, and it may influence the evolution (e.g., Ablimit 2019; Ginzburg \& Quataert 2020)

In this kind of evolution pathway, we find that actually there are some He
stars which can still have masses larger than $M_{\rm Ch}$ after the recycled process.
These super-$M_{\rm Ch}$ He stars may explode as type Ib supernovae
and collapse to be NSs (e.g. Yoon et al. 2010). Thus,
there is possibility of formation of double NS systems, which is one of the good
candidates for gravitational wave sources. In this work, we only
focused on the formation of NS + WD systems.
The predicted orbital period and WD mass distributions of
the binary pulsar systems given in Figure 4 are also in good
agreement with those in previous works
(Tauris \& Savonije 1999;
Istrate et al. 2014; Liu et al. 2018),
and different regions are the correspondence to the different
progenitor evolutionary pathways mentioned in above sections.
Comparing with observed pulsar - WD systems,
evolutions of ONeMg WD + He stars binaries tend to produce MSPs with
intermediate-mass WD companions and intermediate orbital periods (e.g., Liu et al. 2018). As shown in Figure 4,
MSPs with lower mass WD companions (down to $\sim 0.3 M_\odot$) in orbits with narrow range between 0.08 day and $\sim$ 100 days can
be formed with the magnetic confinement model comparing to that of non-magnetic case
(the lowest mass of the WD companion is around 0.92 $M_\odot$, and the orbital periods range from 0.1 to few hundred days),
because initial conditions and mass transfer phase are changed due to the magnetic confinement as discussed in above sections.
However, the contribution of WD + He star binaries with the magnetic confinement case to MSPs - WD systems is roughly the same as that
of the non-magnetic case, both of them have rates of few$\times10^{-4}\,{\rm yr}^{-1}$ which is consistent with previous works (e.g., Hurley et al. 2010).

\section{Conclusions}

A number of physical elements and processes, such as uncertainties in the nature of newborn NS through AIC,
the kick velocity, the accretion discs, the effect of evaporation, the
rotation, metallicity, CE evolution, and physical response at high mass-transfer rates, may cause significant
discrepancy in results of ONeMg WD + He star binary evolution. In this work, we focus on the
possible role and effect of the WD's strong magnetic field in the evolution of ONeMg WD + He star binary.
In the pre-AIC evolution, the He accretion phase may be influenced by the WD's magnetic field.
Our detailed evolution model results demonstrate the effective accretion under magnetic confinement that makes
it possible for the WDs to steadily accumulate mass with relatively low mass transfer rate by avoiding strong He
flashes which would be experienced in non- or low-magnetic models. This makes differences between the outcomes of WD + He star
binaries with and without magnetic confinement model, but the magnetism of WDs does not cause dramatic
changes in the results derived from WD + He star
binaries as progenitors of AIC or SNe Ia. Note that the accretion phase of CO WD + He star
binaries is similar to that of ONeMg WD + He star binaries, thus conclusions of this work can be
applied for the generation of SNe Ia: the magnetic confinement model drives the orbital periods and He star masses of CO WD + He star
binaries to the lower positions in the progenitor space of SNe Ia (survived companions after the SNe or in the SN remnants would be
 harder to detect), and the magnetism in the CO WD + He star
binary can not significantly alter the contribution of this channel to the SNe Ia.
Derived rates are still lower than the observationally inferred rate of SNe Ia.

Because accreted matter can make the WD rotate faster,
we assume a fast rotating/non(or low)-magnetic ONeMg WDs may form MSPs,
and a fast rotating with the spin period of seconds to milliseconds /highly magnetized (as the level of polars) ONeMg WDs may produce magnetars via AIC.
Our BPS results show that the magnetic confinement model does not change the contribution of ONeMg + He star binary channel to the AIC.
Predicted birthrate of all MSPs formed through AIC is $1.49\times10^{-4}\,{\rm yr}^{-1}$,
and Galactic rate of magnetars which could be formed via AIC of sufficiently magnetized WDs is $0.34\times10^{-4}\,{\rm yr}^{-1}$.
The confined accretion lower the limit for the
initial mass of the He donor star which let the WD reach the Chandrasekhar limit mass,
and it means that the mass of the newborn MSP's companion star after the
AIC can be lower (at least $\sim 0.4 M_\odot$) than that of non-magnetic case.

\section*{Acknowledgements}

I thank the referee for the useful comments that improve this work.
This work is supported by NSFs.

\textbf{Data Availability:} The data underlying this article
will be shared on reasonable request to the corresponding author.

%%%%%%%%%%%%%%%%%%%%%%%%%%%%%%%%%%%%%%%%%%%%%%%%%%



%\bibliographystyle{mnras}


Abdusalam, K., Ablimit, I., Hashim, P., L$\ddot{\rm u}$, G.-L. et al. 2020, ApJ, 902, 125

Ablimit, I. \& Li, X.-D., 2015, ApJ, 800, 98

Ablimit, I., Maeda, K. \& Li, X.-D., 2016, ApJ, 826, 53

Ablimit, I., \& Maeda, K., 2018, ApJ, 866, 151

Ablimit, I., \& Maeda, K. 2019a, ApJ, 871, 31

Ablimit, I., \& Maeda, K. 2019b, ApJ, 885, 99

Ablimit, I. 2019, ApJ, 881, 72

Ablimit I., 2021, PASP, 133, 074201

Ablimit, I., Podsiadlowski, P., Hirai, R. \& Wicker, J. 2021, eprint arXiv:2108.08430

Alpar, M. A., Cheng, A. F., Ruderman, M. A., \& Shaham, J. 1982, Nature,
300, 728

Angel, J. R. P. 1978, ARA\&A, 16, 487

Benvenuto, O. G., De Vito, M. A., \& Horvath, J. E. 2014, ApJL, 786, L7

Bhattacharya, D., \& van den Heuvel, E. P. J. 1991, PhR, 203, 1

Bildsten, L., Shen, K. J., Weinberg, N. N., \& Nelemans, G. 2007, ApJL,
662, L95

Brooks, J., Bildsten, L., Marchant, P., \& Paxton, B. 2016, ApJ, 821, 28

Brooks, J., Schwab, J., Bildsten, L., Quataert, E.,  \& Paxton, B. 2017, ApJ, 843,
151

Chanmugam, G., \& Brecher, K. 1987, Natur, 329, 696

Connelley, M. S., Reipurth, B. \& Tokunaga, A. T. 2008, AJ, 135, 2526

Cropper, M. 1990, SSRv, 54, 195

Cumming, A. 2002, MNRAS, 333, 589

Deloye, C. J., Taam, R. E., Winisdoerffer, C., \& Chabrier, G. 2007, MNRAS, 381, 525

Duncan, R. C., \& Thompson, C. 1992, ApJL, 392, L9

Dubus, G., Lasota, J.-P., Hameury, J.-M., \& Charles, P. 1999, MNRAS,
303, 139

Eggleton, P. P. 1983, ApJ, 268, 368

Fabian, A. C., Pringle, J. E., Rees, M. J. \& Whelan, J. A. J. 1977, MNRAS, 179, 9

Ferrario, L. 2012, MNRAS, 426, 2500

Ferrario, L. \& Wickramasinghe, D. 2007, MNRAS, 375, 1009

Ferrario, L., de Martino, D., \& Gaensicke, B. T. 2015, SSRv, 191, 111F
%\bibitem[Fujimoto{1982}]{f82b}

Frank, J., King, A., \& Raine, D. J. 2002, Accretion Power
in Astrophysics, by Juhan Frank and Andrew King and
Derek Raine, pp. 398. ISBN 0521620538. Cambridge,
UK: Cambridge University Press, February 2002., 398

Freire, P.C.C. \& Tauris, T.M. 2014, MNRAS 438(1), L86

Ghosh P. \& Lamb F. K., 1979, ApJ, 232, 259

Hachisu, I., Kato, M., \& Nomoto, K. 1996, ApJL, 470, L97

Hameury, J.-M., King, A. R., \& Lasota, J.-P. 1986,
MNRAS, 218, 695

Hillman, Y. 2021, preprint, arXiv:2106.13006

Hurley, J. R., Tout, C. A. \& Pols, O. R. 2002, MNRAS, 329, 89

Hurley, J. R., Tout, C. A., Wichramasinghe, D. T., Ferrario, L., \& Kiel, P. D.
2010, MNRAS, 402, 1437

Iben, I., \& Tutukov, A. 1991, ApJ, 370, 615

Istrate, A., Tauris, T., \& Langer, N. 2014, A\&A, 571, A45

Kato, M., Hachisu, I., Kiyota, S., \& Saio, H. 2008, ApJ, 684, 1366

Kato, M., \& Hachisu, I. 2004, ApJL, 613, L129

Kahabka, P. 1995, ASP Conference Series, Vol. 85

Kepler, S. O., Pelisoli, I., Jordan, S., et al. 2013, MNRAS, 429, 2934

Kepler, S. O., Pelisoli, I., Koester, D., et al. 2015, MNRAS, 446, 4078

Klencki, J., Nelemans, G., Istrate, A. G. et al. 2021, A\&A, 645, 54

King, A. R. \& Shaviv, G. 1984, MNRAS, 211, 883

King, A. R., MNRAS, 1993, 261, 144

Kirsten, F., Marcote, B., Nimmo, K. et al. 2021, eprint arXiv:2105.11445

Kulkarni, S. R., \& Narayan, R. 1988, ApJ, 335, 755

Landau, L. D., \& Lifshitz, E. M. 1975, The Classical Theory of Fields (4th ed.;
Oxford: Pergamon)

Levin, L., Bailes, M., Barsdell, R. B., et al. 2013, MNRAS, 434, 1387

Liu, D., Wang, B., Chen, W., Zuo, Z. \& Han, Z. 2018, MNRAS, 477, 384

Livio, M. 1983, A\&A, 121, L7

Manchester, R. N., Hobbs, G. B., Teoh, A. \& Hobbs, M. 2005, AJ, 129, 1993

Michel, F. C. 1987, Natur, 329, 310

Nomoto, K. 1982, ApJ, 253, 798

Nomoto, K., \& Kondo, Y. 1991, ApJL, 367, L19

Olausen, S. A. \& Kaspi, V. M. 2014, ApJS, 212, 6

Osborne, et al. 2001, A\&A, 378, 800

Paxton, B., Marchant, P., Schwab, J., et al. 2015, ApJS, 220, 15

Pylyser, E., \& Savonije, G. J. 1988, A\&A, 191, 57

Pylyser, E. H. P., \& Savonije, G. J. 1989, A\&A, 208, 52

Rappaport, S., Verbunt, F., \& Joss, P. C. 1983, ApJ, 275, 713

Ritter, H. 1988, A\&A, 202, 93

Roberts, M. S. E. 2013, in IAU Symp. 291, Neutron Stars and Pulsars:
Challenges and Opportunities after 80 Years, ed. J. van Leeuwen
(Cambridge: Cambridge Univ. Press), 127

Ruiter, A. J., Ferrario, L., Belczynski, K. et al. 2019, MNRAS, 484, 689

Schmidt, G. D., Hoard, D. W., Szkody, P., Melia, F., Honeycutt, R. K., \& Wagner, R. M. 1999, ApJ, 525, 407

Schwope, A., Buckley, D. A.H., Malyali, A. et al. 2021, eprint arXiv:2106.14540

Shen, K. J. \& Bildsten, L. 2009, ApJ, 699, 1365

Sokoloski, J. L., \& Beldstin, L. 1999, ApJ, 517, 919

Stroeer, A., \& Nelemans, G. 2009, MNRAS, 400, L24

Tauris, T. M., Sanyal, D., Yoon, S.-C., \& Langer, N. 2013, A\&A, 558, A39

Tauris, T. M., \& Savonije, G. J. 1999, A\&A, 350, 928

Townsley, D. M., Moore, K., Bildsten, L., 2012, ApJ, 755, 4

van den Heuvel, E. P. J., \& van Paradijs, J. 1988, Natur, 334, 227

van den Heuvel E. P. J. \& Bitzaraki O., 1995, A\&A, 297, L41

Verbunt, F., \& Zwaan, C. 1981, A\&A, 100, L7

Verbunt, F., Wijers, R. A. M. J., \& Burn, H. M. G. 1990, A\&A, 234, 195

Wang, B. \& Liu, D. 2020, Research in Astronomy and Astrophysics, 20, 135

Wang, C., Jia, K., \& Li, X.-D. 2016, RAA, 16h, 9

Wickramasinghe, D. T. \& Ferrario, L. 2000, PASP, 112, 873

Wickramasinghe, D. 2014, European Physical Journal Web of Conferences, 64,
03001

Willems, B., \& Kolb, U. 2004, A\&A, 419, 1057

Woosley, S. E., \& Kasen, D. 2011, ApJ, 734, 38

Woosley, S. E., \& Weaver, T. A. 1994, ApJ, 423, 371

Yoon S.-C., Langer N., 2005, A\&A, 435, 967

Yoon S. C., Woosley S. E., \& Langer N., 2010, ApJ, 725, 940

\clearpage

\begin{figure*}
\centering
\includegraphics[totalheight=4.5in,width=3.5in]{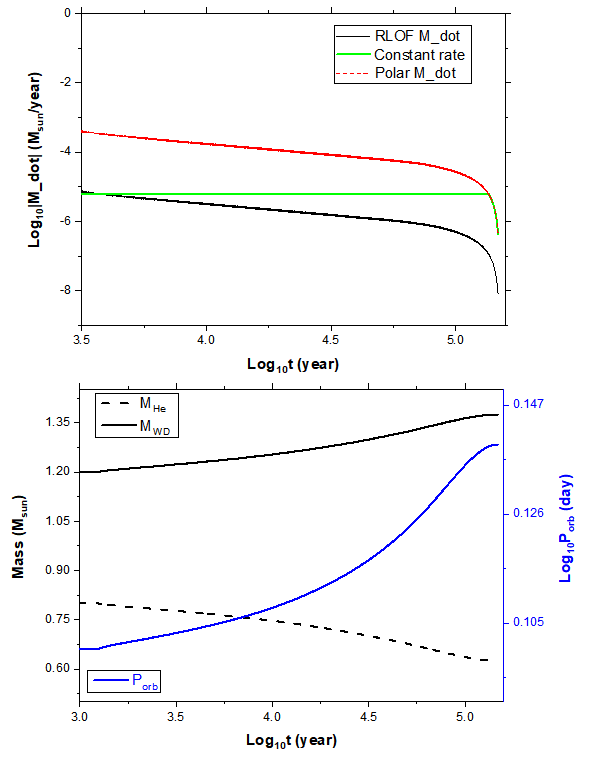}
\caption{. One typical example for pre-AIC evolutions of ONeMg WD binaries under the magnetic confinement model.
The initial masses of the WD and He donor star are 1.2 $M_\odot$ and 0.8 $M_\odot$, respectively.
The initial orbital period is 0.1 day. The upper panel shows evolutions of the RLOF mass-transfer
rate (${\dot{M}}$) and polar mass-transfer rate ($\dot{M}_{\rm{p}}$) with time. The green line is the constant accretion rate that is set when the accretion rate exceeds $10^{-6}\,\rm{M_\odot\,yr^{-1}}$. The lower panel demonstrates changes of the He star mass ($M_{\rm He}$), WD mass ($M_{\rm WD}$) and orbital period (see right axis for $P_{\rm orb}$) with time.}
\label{fig:1}
\end{figure*}

\clearpage

\begin{figure*}
\centering
\includegraphics[totalheight=4.5in,width=3.2in]{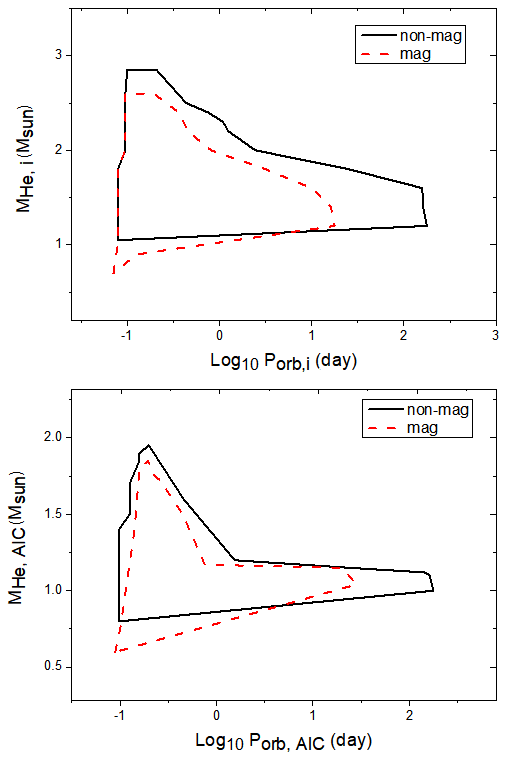}
\caption{Upper panel: Initial parameter space of ONeMg WD $+$ He star binaries which lead to
AIC (it would be SNe Ia if WDs have CO cores) in the initial orbital period ($P_{\rm orb,i}$)-initial secondary mass ($M_{\rm He,i}$) plane. The solid black line is for the non magnetic case (non-mag), while the red dashed line shows results under the magnetic confinement case (mag). The initial WD masses are 1.2 $M_\odot$ for both cases. Lower panel: Evolution results from upper panel, and it shows the distribution of orbital period and He star mass just after the AIC occurs.}
\label{fig:1}
\end{figure*}

\clearpage

\begin{figure*}
\centering
\includegraphics[totalheight=3.5in,width=4.5in]{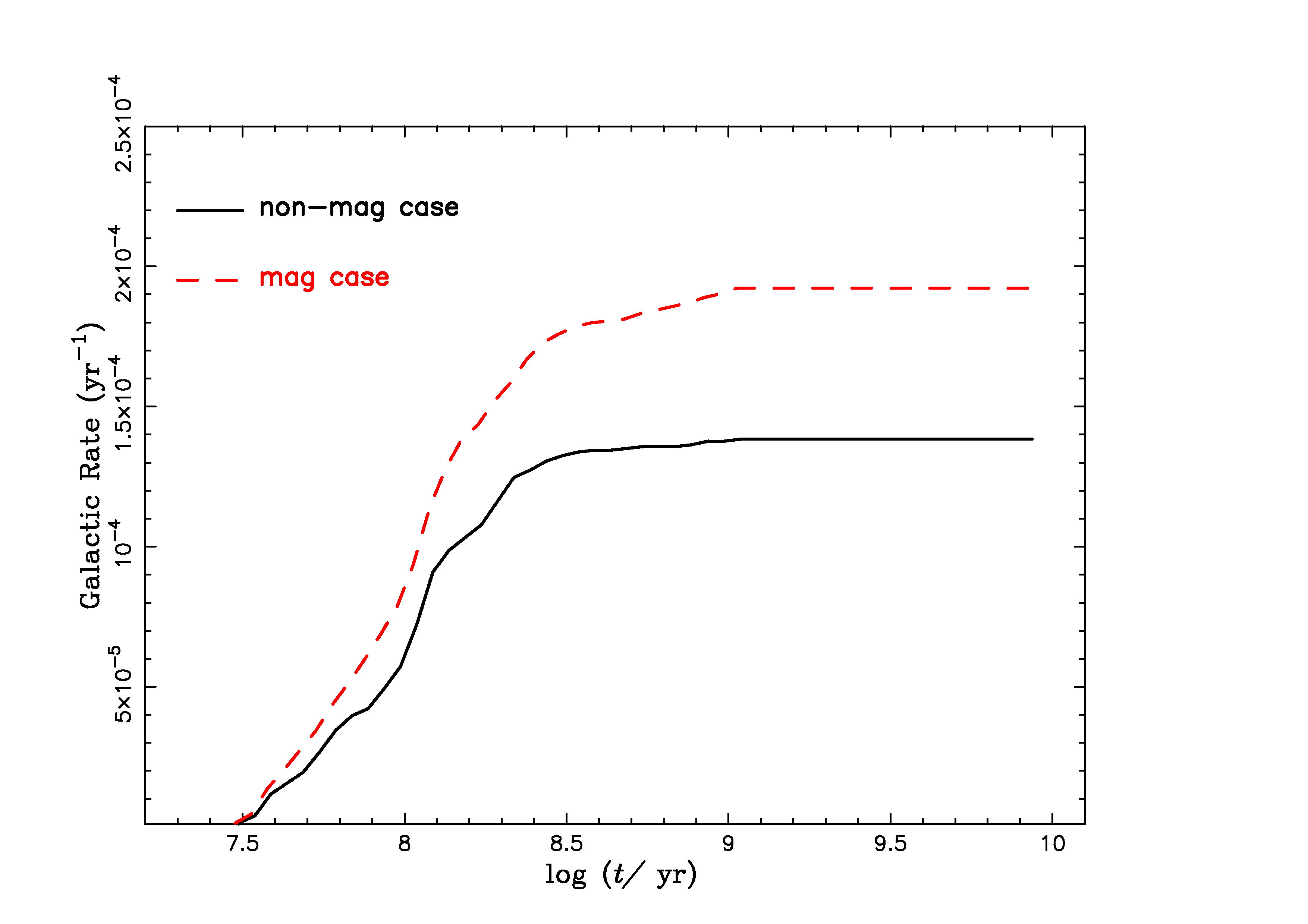}
\caption{The simulated AIC birthrate evolution with a constant SFR for non-magnetic and magnetic cases (see text for more discussions).
}
\label{fig:1}
\end{figure*}

\clearpage

\begin{figure*}
\centering
\includegraphics[totalheight=3.5in,width=4.5in]{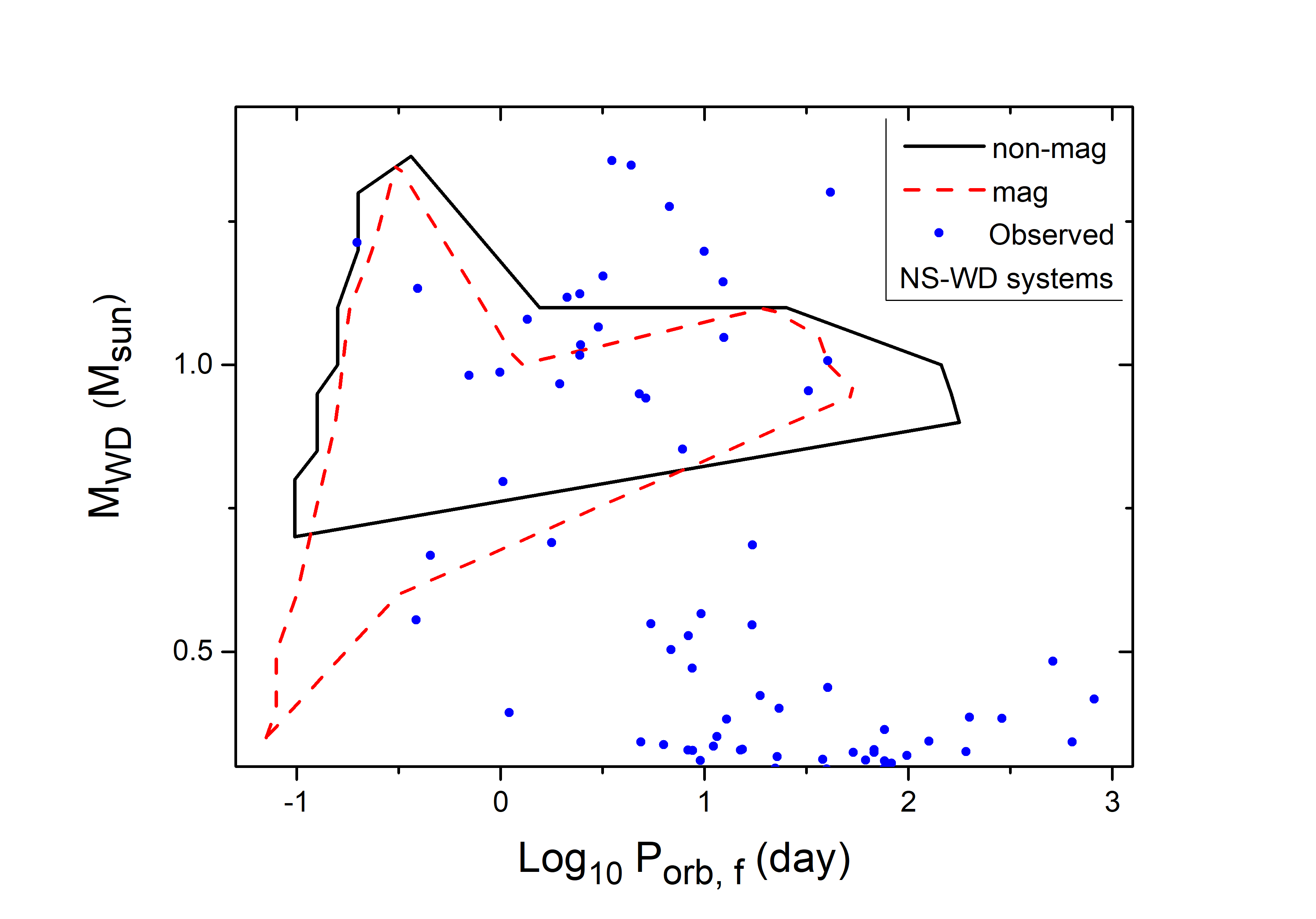}
\caption{WD mass distributions VS. Orbital periods for
MSP-WD systems evolved from binaries of Figure 2. The blue circles represent the observed NS-WD binaries taken from the
ATNF Pulsar Catalogue (Manchester et al. 2005; see
http://www.atnf.csiro.au/research/pulsar/psrcat).
Note that in this work we concentrate on the companion WD mass and orbital period rather than mass of NS.}
\label{fig:1}
\end{figure*}

\clearpage

%%%%%%%%%%%%%%%%%%%%%%%%%%%%%%%%%%%%%%%%%%%%%%%%%%

%%%%%%%%%%%%%%%%% APPENDICES %%%%%%%%%%%%%%%%%%%%%

%\appendix



%%%%%%%%%%%%%%%%%%%%%%%%%%%%%%%%%%%%%%%%%%%%%%%%%%

% Don't change these lines
%\bsp	% typesetting comment
\label{lastpage}
\end{document}